\documentclass[twocolumn,nopacs,floatfix,amsmath,nofootinbib,amssymb,preprintnumbers]{revtex4}
\usepackage{graphicx,color,dcolumn,booktabs,bm}
\usepackage{txfonts}
\usepackage{slashed}
\usepackage[colorlinks,
            citecolor=blue,
            anchorcolor=red,
            menucolor=red,
            linkcolor=red,
            filecolor=red,
            runcolor=red,
            urlcolor=blue,
            frenchlinks=red]{hyperref}

\begin{document}
\title{Pion photoproduction off nucleon with Hamiltonian effective field theory}
\author{Dan Guo}
\author{Zhan-Wei Liu}\email[Corresponding author: ]{liuzhanwei@lzu.edu.cn}

\affiliation{
$^1$School of Physical Science and Technology, Lanzhou University, Lanzhou 730000, China\\
$^2$Research Center for Hadron and CSR Physics, Lanzhou University and Institute of Modern Physics of CAS, Lanzhou 730000, China\\
$^3$Lanzhou Center for Theoretical Physics, Key Laboratory of Theoretical Physics of Gansu Province, and Frontiers Science Center for Rare Isotopes, Lanzhou University, Lanzhou 730000, China}

\begin{abstract}
We analyze the $\gamma N\to\pi N$ process in the negative parity channel with Hamiltonian effective field theory which is successful in studying the nucleon resonances associated with  the meson-nucleon scattering and lattice QCD simulations. We provide the scattering amplitude of pion photoproduction and extract the electric dipole amplitudes $E_{0+}$. The bare $N^*(1535)$ is important for improving the consistency between our results and the experimental data.

\end{abstract}

\maketitle
\section{Introduction}\label{sec1}

An important challenge in hadron physics is to understand the nature of excited nucleon resonances. These excitations are coupled to some meson-baryon and $\gamma N$ channels \cite{Crede:2013kia,Capstick:2000qj}. Extensive data are accumulated from photoproduction, electroproduction, and meson-nucleon scattering for decades of efforts, which can help us disclose the structure of these resonances and furthermore the properties of QCD in the nonperturbative region.

The $\pi N\to\pi N$ scattering is a powerful measured process to study the nucleon excitations \cite{Klempt:2009pi,Crede:2013kia}, and it can determine the baryon masses, widths, pole positions and decay branching ratios \cite{ParticleDataGroup:2020ssz}. The first negative-parity nucleon $N^*(1535)$ is closely related to this process and has been discussed in many works \cite{Doring:2009yv,Zou:2010tc,An:2008xk,Inoue:2001ip,Xiao:2015gra,Nakayama:2008tg,Matsuyama:2006rp,Kamano:2013iva,Eichmann:2018ytt}. Its mass is larger than the Roper state $N^*(1440)$ with positive parity, which is controversial with the prediction of conventional quark model \cite{Capstick:1986ter}.
Thus the pentaquark component is expected in $N^*(1535)$, coupling with the conventional triquark kernel. Moreover, it is also suggested that the resonance may be dynamically generated from meson-nucleon scattering \cite{Kaiser:1995cy,Inoue:2001ip}.

The lattice QCD simulation, from the first principle of QCD, has achieved remarkable successes. It can reproduce the masses and other properties of hadrons on unphysical pion masses in finite volumes \cite{Fodor:2012gf,Leinweber:2015kyz,Kamleh:2017lye}.
The well-known L\"{u}scher formulation and extended models have been used to extract phase shifts from lattice QCD simulation \cite{Luscher:1985dn,Luscher:1986pf}. There are also some other phenomenological approaches which are set in the finite volumes and used to extract the useful physical information from the lattice QCD calculations.

Hamiltonian effective field theory (HEFT) also enables an interpretation linking the experimental data with the lattice QCD energy levels in a consistent way \cite{Liu:2015ktc,Liu:2016uzk,Liu:2016wxq,Wu:2017qve,Wu:2016ixr,Li:2019qvh,Liu:2020foc,Li:2021mob}. Additionally, one can use the eigenvectors of the discretized Hamiltonian to analyze the structures of finite-volume states, which can further predict what the energy levels can be possibly observed with different interpolation operators in lattice QCD.

In another aspect, photoproduction of meson is also an excellent tool to extract information about the mechanism of the strong interaction at low energies \cite{Xing:2018axn,Xie:2013mua,Xiao:2015gra,Zhong:2011ti,Zhong:2011ht,Sibirtsev:2005ns,An:2008xk,Ajaka:2006bn,Zhao:2003gs,Zhao:2002id,Raya:2021pyr,Wang:2022pbr}.
It is helpful to determine the basic properties of nucleon resonances, such as the spin, pole positions, beam asymmetries, anomalous magnetic moments, and electromagnetic couplings  \cite{Workman:2012jf,Kamano:2016bgm,Krusche:2003ik,Sandorfi:2010uv,Nakayama:2008tg,Huang:2011as,Ronchen:2014cna,Kim:2021adl,Wu:2012wta}. The electromagnetic properties of hadrons, such as magnetic moments, are essential for constructing the picture of hadrons and widely discussed with various approaches \cite{Li:2021ryu,Li:2017vmq,Li:2017pxa,Li:2016ezv,Li:2017cfz,Wang:2018xoc,Meng:2018gan,Shanahan:2014cga,Leinweber:1990dv,Segovia:2015hra,Wilson:2011aa,Zou:2005xy,Liu:2005pm}.  In addition, the relevant studies are the necessities for the photon-nucleus investigation which can detect the structure and electromagnetic properties of nuclei. For example, meson photoproductions on $^4\mathrm{He}$ are investigated based on the description of scattering on nucleon \cite{Kim:2021adl,Wu:2012wta}.

At low energies, the elegant chiral perturbation theory is successful in exploring photoproduction processes $\gamma N\to\pi N$, concerning the near threshold region and low partial waves, which often give a precise accord in this region \cite{Ma:2020hpe,GuerreroNavarro:2019fqb,Bernard:1991rt,Bernard:1994gm,Bernard:2007zu,Hilt:2013uf}. References \cite{Ma:2020hpe,Cao:2021kvs} studied $S_{11}$ partial-wave multipole amplitude $E_0^+$ and pion electroproduction process from threshold to below the $\Delta(1232)$ energy region. The K-matrix and its variations are widely used to reproduce experimental data of photoproduction process \cite{Arndt:2002xv,Shklyar:2004ba,Penner:2002md,Usov:2005wy,Shyam:2008fr,Drechsel:1998hk,Drechsel:2007if,Tiator:2011pw}. The $\gamma N\to\pi N$ and $\pi N\to\pi N$ reactions are simultaneously analyzed with $N^*(1535)$ and $N^*(1650)$ both taken into account in Ref. \cite{Doring:2009uc}.

Dynamical coupled-channel models, like the EBAC model \cite{Matsuyama:2006rp,Kamano:2013iva,Kamano:2016bgm}, $\mathrm{J\ddot{u}lich}$ model \cite{Doring:2009uc,Doring:2013glu,Doring:2009yv} and other approaches \cite{Pascalutsa:2004pk,Fuda:2003pd,Chen:2007cy}, are proposed to give a comprehensive view about plenty of partial waves with energy up to 2 GeV. Usually the effective Lagrangians are employed to obtain the Bethe-Salpeter equation kernel, and these models can well interpret the data from SAID partial-wave analysis group \cite{gws}.

The photoproduction can ideally inspect the potential meson-nucleon configuration in nucleon resonances since the electromagnetic properties of the nucleon are measured pretty well. Thus, the comprehensive analysis, combining empirical meson-nucleon scattering, the lattice QCD simulation, and photoproduction on nucleon, can be an ideal platform to promote our understanding of the properties of nucleon resonances.

HEFT uses the potentials as the bridge to connect the infinite-volume and finite-volume results. By solving the dynamical equations like Bethe-Salpeter equation with the potentials, one obtains scattering T matrix and thus other information. We solve the eigenstates and eigenvalues of finite-volume matrix Hamiltonian with discretizing the potentials, which can be used to analyze the lattice QCD simulations. HEFT has been successful in dealing with the systems involving $N^*(1535)$, $N^*(1440)$, $\Lambda(1405)$, kaonic deuteron, and so on \cite{Liu:2015ktc,Liu:2016uzk,Liu:2016wxq,Wu:2017qve,Wu:2016ixr,Li:2019qvh,Liu:2020foc,Li:2021mob}.

In this work we use HEFT to further analyze the $\gamma N\to \pi N$ processes in the negative parity channels based on our previous work in which the pion-nucleon scattering and lattice QCD results have been investigated. With the well-performed analysis of $\pi N\to\pi N$ scattering by HEFT \cite{Liu:2015ktc}, a set of couplings about meson and baryons has been obtained. Having been examined by the lattice QCD results, this set of parameters will not be adjusted in this work. Combining the data of pion photoproduction, we further discuss the structure of $N^*(1535)$ and the electromagnetic properties and strong interactions of relevant hadrons.

We organize this work as follows. In Sec. \ref{sec:Frame} we extend HEFT to deal with the $\gamma N\to \pi N$ scattering T matrix and corresponding multipole amplitudes. The numerical results and discussion are given in Sec. \ref{sec2}, followed by a brief summary in Sec. \ref{summary}.

\section{Framework}\label{sec:Frame}

We present the scattering amplitudes of $\gamma N\to\pi N$ and extract the corresponding electromagnetic multipole amplitudes in this section. The scattering amplitudes are split into two parts. One is the pure electromagnetic potential without considering the finite state interactions (FSI), and the other is FSI correction.

We focus on the finite state $\pi N$ in $S_{11}$ partial wave at low energies, and the coupled channel effects from $\eta N$ are also included. The electromagnetic amplitudes of $\gamma N\overset{\rm EM}{\to}\,\pi N/\eta N$ are provided in Sec. \ref{sec:EM_amp} and the corresponding potentials are listed in Sec. \ref{sec:EM_pttl}. The FSI corrections are dealt within HEFT in Sec. \ref{sec:FSI}. We combine these two contributions $\gamma N\overset{\rm EM}{\to}\,\pi N/\eta N\,\overset{\rm FSI}{\to}\pi N$ to obtain the final results in Sec. \ref{sec:Multipole}.

\subsection{Electromagnetic amplitudes without FSI}\label{sec:EM_amp}

\begin{figure*}[htbp]
  \centering
  \includegraphics[width=\textwidth]{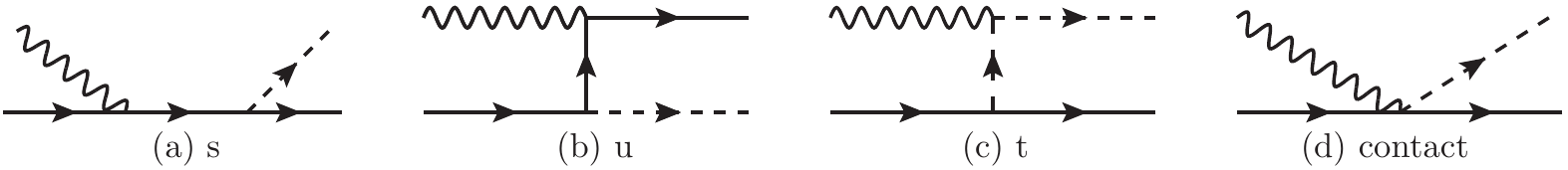}
  \caption{Tree-level diagrams for pure electromagnetic amplitude of $\gamma N\to \pi N/\eta N$ process without FSI: (a) s channel, (b) u channel, (c) t channel (exchanging $\pi$ and $\rho$), (d) the contact term. The solid, wiggly, and dashed lines represent the baryons, photons, and mesons, respectively.}\label{sut}
\end{figure*}

Near threshold the nucleon resonances do not affect the $\gamma N\to \pi N$ process very much. The Lagrangians for $\gamma N\overset{\rm EM}{\to}\,\pi N$ can be written as \cite{Matsuyama:2006rp}
\begin{eqnarray}
  \mathcal{L}_{\pi NN}&=&-\frac{f_{\pi NN}}{m_\pi}\bar{N}\gamma_\mu\gamma_5 \vec{\tau} \cdot \partial^\mu\vec{\pi} N, \\
  \mathcal{L}_{\gamma NN}&=&e\bar{N}\left[ \hat{e}_N\gamma^\mu A_\mu + \frac{\hat{\kappa}_N}{4m_N}\sigma^{\mu\nu} F_{\mu\nu}\right] N , \\
  \mathcal{L}_{\gamma\pi\pi}&=&e\left[ \vec{\pi}\times\partial^\mu\vec{\pi} \right]_3 A_\mu, \\
  \mathcal{L}_{\gamma N\pi N}&=&e\frac{f_{\pi NN}}{m_\pi}\bar{N} \gamma^\mu\gamma_5 \left[ \vec{\tau}\,\times\vec\pi \right]_3 N A_\mu, \\
  \mathcal{L}_{\gamma \rho\pi}&=&e\frac{g_{\gamma \rho\pi}}{m_\pi}\varepsilon^{\mu\nu\alpha\beta} \vec{\pi}\cdot \partial_\mu\vec{\rho}_\nu\partial_\alpha A_\beta, \\
  \mathcal{L}_{\rho NN}&=&g_{\rho NN}\bar{N} \left[ \gamma_\mu - \frac{\kappa_\rho}{2m_N} \sigma_{\mu\nu}\partial^\nu\right]\vec{\rho}\,^\mu\cdot\frac{\vec{\tau}}{2} N ,
\end{eqnarray}
where $\hat e_N\equiv {\rm diag}\{+1,0\}$, the electromagnetic field tensor $F_{\mu\nu}=\partial_\mu A_\nu - \partial_\nu A_\mu$, and the nucleon isospinor $N=(p,n)^T$.  $\hat\kappa_N\equiv {\rm diag}\{\kappa_p,\kappa_n\}$, where $\kappa_p=\mu_p-1=1.79$ and $\kappa_n=\mu_n=-1.91$ with $\mu_p$ and $\mu_n$ being the magnetic moments of proton and neutron in units of nuclear magneton $\mu_N$. These well-defined Lagrangians are constrained by various symmetry properties, such as the invariance under isospin, parity and gauge transformation.
We list the relevant Lagrangians for $\gamma N\overset{\rm EM}{\to}\,\eta N$ \cite{Matsuyama:2006rp}
\begin{eqnarray}
  \mathcal{L}_{\eta NN}&=&-\frac{f_{\eta NN}}{m_\eta}\bar{N} \gamma_\mu\gamma_5 N \partial^\mu \eta, \\
  \mathcal{L}_{\gamma\rho\eta}&=&e\frac{g_{\gamma\rho\eta}}{m_\rho}\varepsilon^{\mu\nu\alpha\beta} \partial_\mu\rho^0_\nu\partial_\alpha A_\beta \eta.
\end{eqnarray}

In this work, the $\pi N$ final state with the total isospin $I$ being $\frac12$ is mainly concerned, and thus for concise expressions we define
\begin{equation}
\hat{\mathcal{M}}_{\pi N}\equiv
\left(\begin{array}{cc}
\hat{\mathcal{M}}_{\pi N, \gamma p}&0\\
0&\hat{\mathcal{M}}_{\pi N, \gamma n}
\end{array}\right),
\end{equation}
where $\hat{\mathcal{M}}_{\pi N, \gamma p}$ specially means the amplitude of $\gamma p\to(\pi N)_{I=\frac{1}{2},I_3=+\frac12}$, and $\hat{\mathcal{M}}_{\pi N, \gamma n}$ refers to that of $\gamma n\to(\pi N)_{I=\frac{1}{2},I_3=-\frac12}$.

With the above Lagragians, we can write out the electromagnetic amplitude of $\gamma(\vec q)\, N(\vec p)\to[\pi(\vec k)\, N(\vec p\,')]_{I=1/2}$ from Fig. \ref{sut}
\begin{eqnarray}\label{M0_piN}
&&\hspace{-1em}
\hat{\mathcal{M}}_{\pi N}^{(0)} =\frac{\sqrt3 e}{m_\pi}\left\{f_{\pi NN}
\left[
+\frac23 \tau^3 \frac{\tilde{\slashed{k}}\gamma_5}{\tilde{k}^2-m^2_\pi+i\epsilon} (\tilde{k}+k) \cdot \varepsilon_\gamma
-\frac23 \tau^3\slashed{\varepsilon}_\gamma \gamma_5
\right.\right.\nonumber \\&&\left.\left.\qquad
 - \slashed{k}\gamma_5 \frac{1}{\slashed{q}+\slashed{p}-m_N+i\epsilon}\Gamma_N
- \Gamma_N\frac{1}{\slashed{p}-\slashed{k}-m_N+i\epsilon} \slashed{k}\gamma_5
 \right]
\right.\nonumber \\&&\left.\qquad
-\frac{g_{\rho NN}g_{\gamma\rho\pi}}{2}
      \frac{\Gamma_\rho}{\tilde{k}^2-m^2_\rho+i\epsilon}\right\}.\qquad
\end{eqnarray}
where $\varepsilon_\gamma$ is the photon polarization vector, $\Gamma_N=\hat{e}_N\slashed{\varepsilon}_\gamma + \frac{\hat{\kappa}_N}{4m_N} [\slashed{q}\slashed{\varepsilon}_\gamma - \slashed{\varepsilon}_\gamma\slashed{q}]$,
$\Gamma_\rho=i\varepsilon_{\mu\nu\alpha\beta}\tilde{k}^\mu[\gamma^\nu + \frac{\kappa_\rho}{4m_N}(\gamma^\nu\tilde{\slashed{k}}-\tilde{\slashed{k}}\gamma^\nu)] q^\alpha \varepsilon^\beta_\gamma$,
and $\tilde{k}^\mu=k^\mu-q^\mu$.

From the diagrams (a), (b), and (c) in Fig. \ref{sut}, the electromagnetic amplitude of $\gamma(\vec q)\, N(\vec p)\to\eta(\vec k)\, N(\vec p\,')$ is
\begin{eqnarray}
&&\hat{\mathcal{M}}_{\eta N}^{(0)} = e\frac{f_{\eta NN}}{m_\pi} \left[ -\slashed{k}\gamma_5 \frac{1}{\slashed{q}+\slashed{p}-m_N+i\epsilon}\Gamma_N
\right.\nonumber \\  &&\left.
- \Gamma_N\frac{1}{\slashed{p}-\slashed{k}-m_N+i\epsilon} \slashed{k}\gamma_5 \right]
- \,e\frac{g_{\rho NN}g_{\gamma\rho\eta}}{m_\eta} \frac{\tau^3}{2}    \frac{\Gamma_\rho}{\tilde{k}^2-m^2_\rho+i\epsilon}.\qquad
\end{eqnarray}

In addition, the contribution from the resonance $N^*(1535)$ should be taken into account especially as the energy approaches the mass since it has the same quantum number with the $S_{11}$ $\pi N$ channel. We use the following Lagrangians \cite{Shyam:2008fr,Hyodo:2006gcx,Nakayama:2008tg,Suh:2018yiu}
\begin{equation}\label{LagGamma}
  \mathcal{L}_{\gamma NN^*}= e\bar N^*  \frac{\hat{g}_{N\gamma N^*}}{4m_N}\gamma_5 \sigma_{\mu\nu}F^{\mu\nu} N +\mathrm{H.c.},
\end{equation}
\begin{equation}\label{LagPiN}
 \mathcal{L}_{\pi NN^*}= \frac{g_{\pi NN^*}}{f_\pi} \bar N^*\gamma^\mu\vec{\tau}\cdot\partial_\mu\vec\pi N + \mathrm{H.c.},
\end{equation}
\begin{equation}\label{LagEtaN}
 \mathcal{L}_{\eta NN^*}= \frac{\sqrt{3}g_{\eta NN^*}}{f_\pi} \bar{N}^*\gamma^\mu\partial_\mu\eta N + \mathrm{H.c.}.
\end{equation}
The electromagnetic amplitudes of $\gamma N \to \pi N/\eta N$ involving $N^*(1535)$ as intermediate state through s and u channels are
\begin{eqnarray}\label{eq:M_Nstar}
&&  \hat{\mathcal{M}}_{\alpha}^{(N^*)}= e \frac{\sqrt{3} g_{\alpha N^*}}{f_\pi} \frac{\hat{g}_{N\gamma N^*}}{4m_N}
 \left\{ \slashed{k} \frac{1}{\slashed{q}+\slashed{p}-m_{N^*}^0+i\epsilon}  \gamma_5(\slashed{q}\slashed{\varepsilon}_\gamma- \slashed{\varepsilon}_\gamma\slashed{q})
\right. \nonumber\\&&\left.\qquad \qquad\quad
+ \gamma_5(\slashed{q}\slashed{\varepsilon}_\gamma- \slashed{\varepsilon}_\gamma\slashed{q}) \frac{1}{\slashed{p}-\slashed{k}-m_{N^*}^0+i\epsilon} \slashed{k} \right\} u(k),
\end{eqnarray}
where $\alpha$ refers to the channel $\pi N$ or $\eta N$, and $m_{N^*}^0$ is the $N^*$ bare mass. The dipole form factor, $u(k)=(1+k^2/\Lambda^2)^{-2}$, regulates the high momentum divergence with cutoff parameter $\Lambda= 0.8$ GeV.
$\hat{g}_{N\gamma N^*}\equiv {\rm diag}\{g_{p\gamma N^{*+}},g_{n\gamma N^{*0}}\}$ is the bare $N^*$-$N$-$\gamma$ coupling.

One obtains the final scattering amplitude $\hat{\mathcal{M}}_{\alpha}$ by summing the above two contributions $\hat{\mathcal{M}}_{\alpha}=\hat{\mathcal{M}}_{\alpha}^{(0)}+\hat{\mathcal{M}}_{\alpha}^{(N^*)}$. In the general case, the Lorentz and gauge invariant amplitude $\hat{\mathcal{M}}_{\alpha;\gamma N}$ can be decomposed into Chew-Goldberger-Low-Nambu (CGLN) amplitudes after preforming the nonrelativistic reduction \cite{Chew:1957tf},
\begin{eqnarray}\label{CGLN}
\hat{\mathcal{M}}_{\alpha,\gamma N}&=&\vec{\sigma}\cdot\vec{\varepsilon}\,t_{\alpha;\gamma N}^{(1)}
  +i\vec{\sigma}\cdot\hat{k}\,\vec{\sigma}\cdot\hat{q}\times\vec{\varepsilon}\,t_{\alpha;\gamma N}^{(2)} \nonumber \\
 &&  + \vec{\sigma}\cdot\hat{q}\,\hat{k}\cdot\vec{\varepsilon}\,t_{\alpha;\gamma N}^{(3)} +
  \vec{\sigma}\cdot\hat{k}\,\hat{k}\cdot\vec{\varepsilon}\,t_{\alpha;\gamma N}^{(4)},
\end{eqnarray}
where $\hat k$ and $\hat q$ mean the unit vectors of $\vec k$ and $\vec q$, respectively. $t_{\alpha;\gamma N}^{(i)}$ is the function of $\vec k$ and $\vec q$ in the center-of-mass reference frame, that is, $t_{\alpha;\gamma N}^{(i)}=t_{\alpha;\gamma N}^{(i)}(\vec k,\vec q)$.

For the process $\gamma(\vec q)\, N(\vec p)\to\pi/\eta(\vec k)\, N(\vec p\,')$, there are $\vec{p}\,'=-\vec{k}$ and $\vec{p}=-\vec{q}$ in the center-of-mass reference frame, and we can choose the direction of photon momentum $\vec q$ along the $z$ axis. The $z$-component of the initial nucleon spin equals to its negative helicity $-\lambda_N$. We show the complete scattering amplitude of $\gamma N\to \pi/\eta\,N$ with explicit spin directions
\begin{eqnarray}\label{explicitAmp}
&&\mathcal M_{\alpha,\gamma N}(s_z^{\prime N},\lambda_N, \lambda_\gamma;\vec{k},\vec{q})
\nonumber \\
&\equiv&
\langle\pi/\eta\,(\vec{k}),\,N(\vec{p}\,',s_z^{\prime N})|\,H^{\rm EM}\,|\gamma(\vec{q},\lambda_\gamma),\,N(\vec{p},-\lambda_N)\rangle
\nonumber \\
&=&
\frac{1}{(2\pi)^3} \frac{1}{\sqrt{2\omega_{\pi/\eta}(k)}}\frac{1}{\sqrt{2|\vec q|}}\,\,\bar{u}_N(s_z^{\prime N})\, \hat{\mathcal{M}}_{\alpha,\gamma N}\, u_N(-\lambda_N),
\qquad
\end{eqnarray}
where $\omega_{\pi/\eta}(k)=\sqrt{m_{\pi/\eta}^2+k^2}$ is the energy of $\pi$ or $\eta$ meson, $u_N$ and $\bar u_N$ are the Pauli spinors for nucleons,
and the photon polarization vector with helicity $\lambda_\gamma=\pm 1$ is expressed as $\varepsilon_\gamma^\pm =\mp\frac{1}{\sqrt{2}}(\hat{x}\pm i\hat{y})$.

\subsection{Electromagnetic potentials without FSI}\label{sec:EM_pttl}

The electromagnetic potentials are directly related to the scattering amplitudes in Eq.  (\ref{explicitAmp}). However, Eq.  (\ref{explicitAmp}) gives the potentials about $|\gamma N\rangle\to |\pi/\eta\,(\vec{k}),\,N(-\vec k,s_z^{\prime N})\rangle$ while we need those about $|\gamma N\rangle\to |\pi/\eta\,N; k, J, J_z, L\rangle$ at last. Here $J$ and $L$ are total angular momentum and the orbital angular momentum of $\pi/\eta\,N$, respectively. For the $z$-component of $J$, there is $J_z=\lambda_\gamma-\lambda_N$.

Before obtaining the potentials of $|\gamma N\rangle\to |\pi/\eta\,N; k, J, J_z, L\rangle$, we first introduce those for $|\gamma N\rangle\to |\pi/\eta\,N; k, J, J_z, \lambda_N'\rangle$ with $\lambda_N'$ being the helicity of the outgoing nucleon
\begin{eqnarray}\label{wignerZ}
&&V_{\alpha,\gamma N}(J,\lambda_N',\lambda_\gamma,\lambda_N;k,q) =
2\pi \int_{-1}^{1}\mathrm{d}(\cos\theta) \; \sum_{s_z^{\prime N}}
\nonumber \\  &&  \qquad
d^{J}_{\lambda_\gamma-\lambda_N,-\lambda_N'}(\theta)
d^{1/2}_{s_z^{\prime N},-\lambda_N'}(\theta)^*
\mathcal M_{\alpha,\gamma N}(s_z^{\prime N},\lambda_N, \lambda_\gamma;\vec{k},\vec{q}),
\qquad
\end{eqnarray}
where $\theta$ is the angle between $\vec k$ and $\vec q$, and $d^{j}_{m,m'}(\theta)$ is the Wigner rotation matrix. About the transformations between different representations one can see Refs. \cite{Jacob:1959at,Richman:1984gh}. With the intermediate potentials in Eq.  (\ref{wignerZ}), we can provide the final potentials we concern
\begin{eqnarray}\label{JLSheliPoten}
V^{JLS;\lambda_\gamma\lambda_N}_{\alpha,\gamma N}(k,q)&=& \sqrt{\frac{2L+1}{2J+1}}\sum_{\lambda_N'}
\langle L,S,0,-\lambda_N' |J,-\lambda_N' \rangle
\nonumber \\  & &\qquad \times
V_{\alpha,\gamma N}(J,\lambda_N',\lambda_\gamma,\lambda_N;k,q).\qquad
\end{eqnarray}
Here $S$ is the total spin of $\pi/\eta\, N$ and equals to $\frac12$ since the spin of $\pi$ or $\eta$ is 0.

We introduce the short notation $V_{\alpha,\gamma N}^{\lambda_\gamma, \lambda_N}$ for the final state $\pi N$ with $J=\frac12, L=0, S=\frac12$
\begin{eqnarray}
V_{\alpha,\gamma N}^{\lambda_\gamma, \lambda_N}(k,q)
&\equiv&
V^{J=\frac12, L=0, S=\frac12;\lambda_\gamma, \lambda_N}_{\alpha,\gamma N}(k,q).
\end{eqnarray}
From Eqs. (\ref{CGLN})-(\ref{JLSheliPoten}) we can obtain the relation
\begin{eqnarray}\label{potenV}
&&V_{\alpha,\gamma N}^{\lambda_\gamma=1, \lambda_N=\frac12}(k,q) = \frac{1}{8\pi^2} \frac{-1}{\sqrt{\omega_{\pi/\eta}(k) q}} \int_{0}^{\pi}\mathrm{d}\theta \; \Big[ 2\sin\theta t_{\alpha,\gamma N}^{(1)} \nonumber \\
&& \qquad + 2\sin\theta\cos\theta t_{\alpha,\gamma N}^{(2)} + 0\cdot t_{\alpha,\gamma N}^{(3)} + \sin^3\theta t_{\alpha,\gamma N}^{(4)} \Big].
\end{eqnarray}

\subsection{FSI within HEFT}\label{sec:FSI}
In our previous work \cite{Liu:2015ktc}, HEFT is successful in studying both $\pi N\to\pi N$ scattering and the related nucleon resonance spectra in finite volume. Hence, the same formalism and parameters are now also adopted this work for the FSI effects. We briefly repeat the relevant formalism in this subsection.

The interacting Hamiltonian includes two parts, $H_I=g+v$. The $g$ describes the vertex interaction between bare state $N^*$ and coupled channel $\alpha$,
\begin{equation}\label{}
  g= \sum_{\alpha}\int\mathrm{d^3}\vec k \,\left\{|\alpha(\vec k)\rangle\, G_\alpha^\dagger(k)\,\langle N^*| + |N^*\rangle\, G_\alpha(k)\,\langle\alpha(\vec k)| \right\},
\end{equation}
where $G_\alpha(k)$ depicts the ordinary S-wave coupling and can be derived from the Lagrangians Eqs. (\ref{LagPiN})-(\ref{LagEtaN})
\begin{equation}\label{piN_vertex}
  G_\alpha(k)= \frac{\sqrt{3} g_{\alpha N^*}}{2\pi f_\pi}\sqrt{\omega_{\pi/\eta}(k)}\,u(k).
\end{equation}
The other interacting part, $v$, is phenomenologically describing direct two-to-two particle interaction by
\begin{equation}\label{}
  v=\sum_{\alpha,\beta}\int \mathrm{d^3}\vec{k}\mathrm{d^3}\vec k' \,|\alpha(\vec k)\rangle\, V^S_{\alpha,\beta}(k,k')\, \langle\beta(\vec{k}')|.
\end{equation}
Here, the separable potentials, $V^S_{\alpha,\beta}(k,k')$, is introduced.
For instance, the potential for $\pi N$ channel direct scattering is
\begin{equation}\label{}
  V^S_{\pi N,\pi N}(k,k')=\frac{g^S_{\pi N}\tilde{u}(k)\tilde{u}(k')}{4\pi^2 f_\pi^2}.
\end{equation}
where $\tilde{u}(k)=u(k)[m_\pi+\omega_\pi(k)]/\omega_\pi(k)$.

Then solving the $\pi N\to\pi N$ coupled-channel Bethe-Salpeter equation, one can get the corresponding partial wave T-matrix,
\begin{eqnarray}\label{hadronT}
&& T_{\pi N,\pi N}(k, k'; E)= V_{\pi N,\pi N}(k, k'; E)+\sum_{\alpha}\int \mathrm{d} k''\, k''^2
\nonumber\\ &&\qquad\times
V_{\alpha,\pi N}(k'',k; E)\frac{1}{E-\omega_\alpha(k'')+i\epsilon}T_{\pi N,\alpha}(k',k''; E),\qquad
\end{eqnarray}
where $\omega_\alpha(k)=\omega_{\pi/\eta} + \omega_{N}=\sqrt{m_{\pi/\eta}^2+k^2}+\sqrt{m_N^2+k^2}$ is the total energy of the channel $\alpha$, and the coupled-channel potential are calculated from interacting Hamiltonian
\begin{equation}
V_{\alpha,\beta}(k, k';E)= G_{\alpha,N^*}^\dag(k)\frac{1}{E-m_{N^*}^0} G_{\beta,N^*}(k')+V_{\alpha,\beta}^S(k, k').
\end{equation}

\subsection{Pion photoproduction off nucleon with FSI}\label{sec:Multipole}

With the electromagnetic potentials $V^{\lambda_\gamma, \lambda_N}_{\alpha,\gamma N}$ in Eq. (\ref{potenV}) and the scattering T-matrix $T_{\beta,\alpha}$ in Eq. (\ref{hadronT}), one can obtain the scattering T-matrix for $\gamma N\to\pi N$ including FSI effects \cite{Matsuyama:2006rp,Kamano:2013iva,Kamano:2016bgm}
\begin{eqnarray}\label{t-tot}
 && T^{\lambda_\gamma, \lambda_N}_{\pi N,\gamma N}(k,q;E)=V^{\lambda_\gamma, \lambda_N}_{\pi N,\gamma N}(k,q)+\sum_{\alpha}\int \mathrm{d} k'\, k'^2
\nonumber \\  &&\qquad \times
 V^{\lambda_\gamma, \lambda_N}_{\alpha,\gamma N}(k', q)\frac{1}{E-\omega_\alpha(k')+i\epsilon} T_{\pi N,\alpha}(k, k'; E).\quad
\end{eqnarray}
Here we mainly concern the outgoing $\pi N$ with the quantum number $S_{11}$, and such T-matrix is related to the electric dipole $E_{0+}$ \cite{Matsuyama:2006rp}
\begin{eqnarray}\label{multipoleEqn}
E_{0+}(E_{\rm cm})&=& \frac{\pi m_N\sqrt{\omega_{\pi}(k_{\rm on})\, q_{\rm on}}}{E_{\rm cm}}
T^{\lambda_\gamma=1, \lambda_N=\frac12}_{\pi N,\gamma N}(k_{\rm on},q_{\rm on};E_{\rm cm}),\qquad
\end{eqnarray}
where $k_{\rm on}$ and $q_{\rm on}$ are the 3-momenta of $\pi N$ and $\gamma N$ corresponding to the total energy $E_{\rm cm}$ in the center-of-mass frame, respectively.

More information of electromagnetic multipoles for meson photoproduction off nucleon can refer to Refs. \cite{Chew:1957tf, Walker:1968xu, Krusche:2003ik}.

\section{Numerical Results and Discussion}\label{sec2}




\begin{figure*}
  \centering
  \begin{tabular}{cc}
\includegraphics[height=6.5cm]{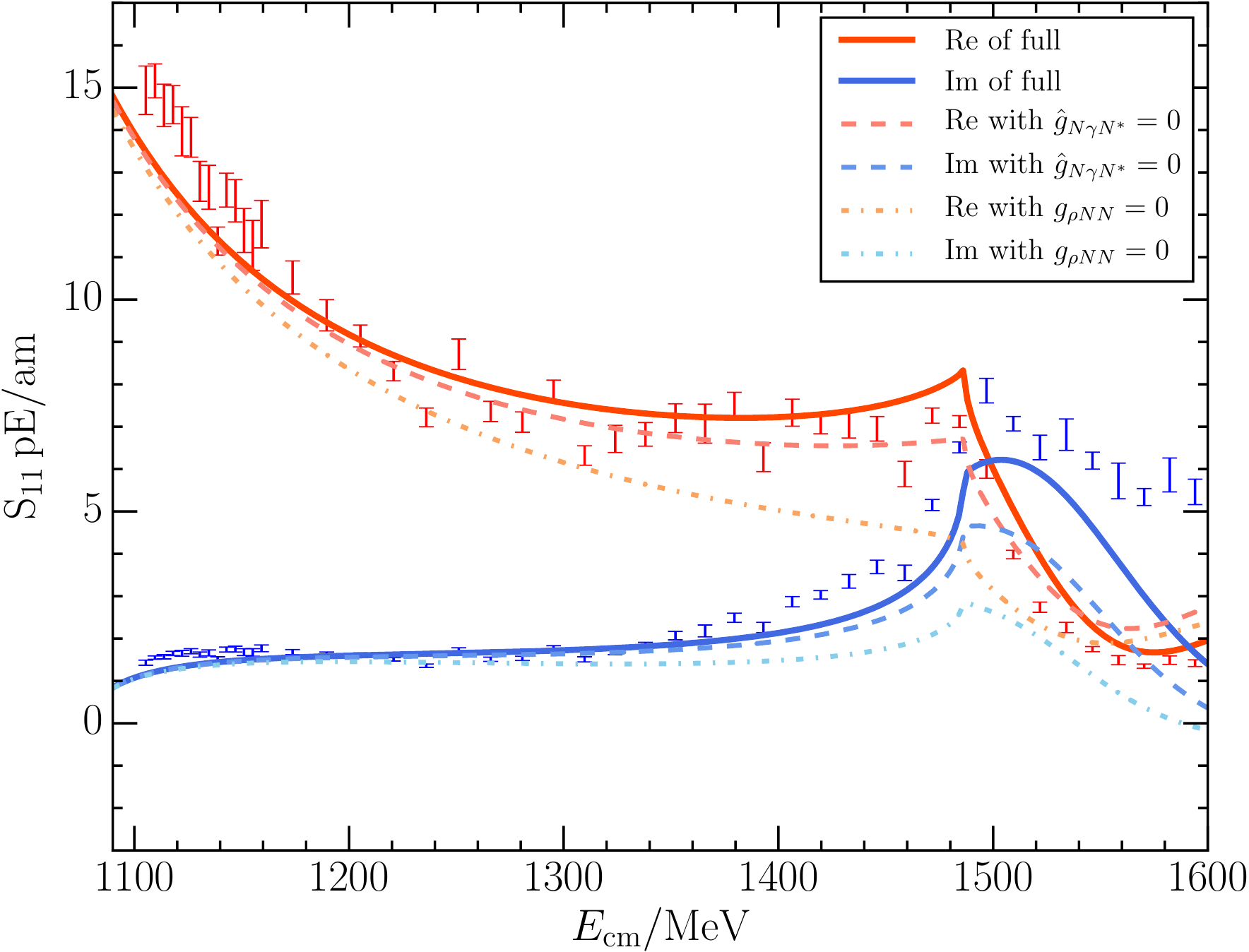}&
\includegraphics[height=6.5cm]{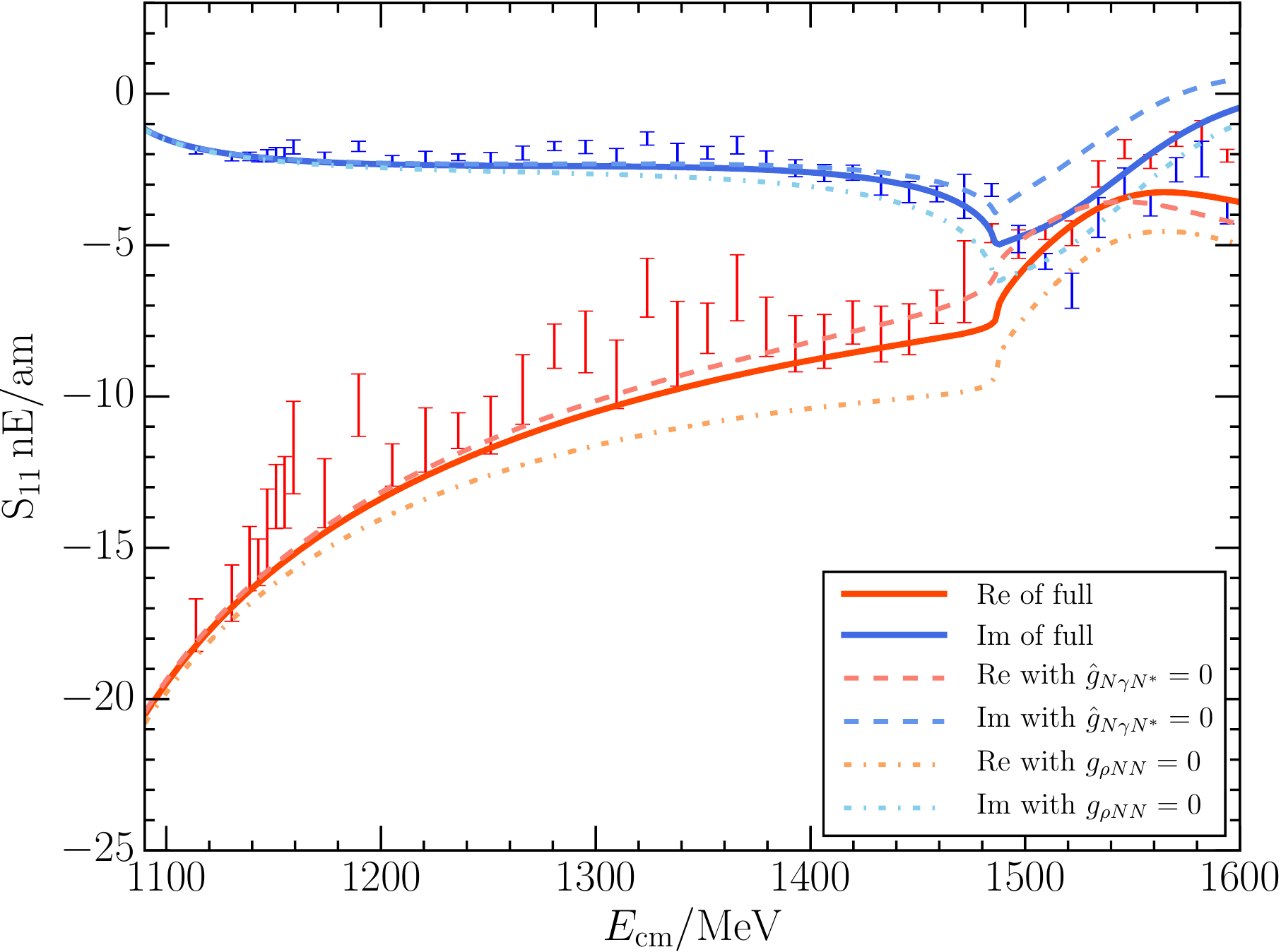}\\
(a)&(b)
  \end{tabular}
  \caption{The electric dipole amplitudes $E_{0+}$ with and without the contributions from the $\rho$ meson or the bare $N^*$ resonance. The solid, dashed and dashed-dotted lines refer to the full, $\hat{g}_{N\gamma N^*}=0$, and $g_{\rho NN}=0$ cases, respectively.}
  \label{multipolerho}
\end{figure*}

We can now obtain the numerical results for the electric dipole amplitudes $E_{0^+}$. We take the couplings and bare resonance mass from our previous HEFT for the FSI part and do not adjust them, that is, $g_{\pi N N^*}=0.186$, $g_{\eta N N^*}=0.185$, $g^S_{\pi N}=-0.0608$, and $m_{N^*}^0=1601$ MeV \cite{Liu:2015ktc}. The common couplings are set as in Refs. \cite{Matsuyama:2006rp,Kamano:2013iva}, $f_{\pi NN}=1.00$, $g_{\gamma\rho\pi}=0.13$, $g_{\gamma\rho\eta}=1.15$, $\kappa_\rho=1.82$, $g_{\rho NN}=6.20$, $f_\pi=92$ MeV. We do not consider the contribution from $\eta'$ mesons in this work, which may be partly absorbed by $f_{\eta NN}$. We fit the experimental data of $E_{0^+}$ and obtain $f_{\eta NN}=-1.42$, $g_{p\gamma N^{*+}}=0.27$ and $g_{n\gamma N^{*0}}=-0.25$. The masses of nucleon and mesons are taken from PDG \cite{ParticleDataGroup:2020ssz}.
We can provide the electric dipole amplitudes $E_{0^+}$ using our approach in Fig. \ref{multipolerho}, along with the  experimental data \cite{gws}.

As shown in Fig. \ref{multipolerho}, the electric dipoles $E_{0+}$  in our framework are compatible with experimental data for both proton and neutron targets up to 1.6 GeV. Our results can describe the data both at low energies close to $\pi N$ threshold and at $N^*$ resonance energies.

The basic contributions $\hat{\mathcal{M}}_{\pi N}^{(0)}$ and $\hat{\mathcal{M}}_{\eta N}^{(0)}$ describe the nonresonant meson-baryon interaction at tree level with the effective Lagrangian approach.
The amplitude $\hat{\mathcal{M}}_{\pi N}^{(0)}$ consists of s- and u-, and t-channel term, and the contact term which depicts the short-range heavy particle exchanging effects and ensure the gauge invariance. Among these terms of tree-level amplitude, the contact term occupies the main share, the t-channel term has moderate contribution, and the s- and u-channel terms are usually smallest. It is that only the antinucleon can propagate in the s channel and u channel because of the parity conservation that makes the corresponding suppression in this case. Since these basic contributions contain none resonance,  they are smooth without bumplike, cusplike or diplike structures.

With the center-of-mass energy $E_{\rm cm}$ increasing, the absolute magnitudes of the contact and the pion-exchanging t-channel terms in $\hat{\mathcal{M}}_{\pi N}^{(0)}$ are continuously declining, the u-channel contribution also declines and changes the sign, and the s-channel part varies little. The $\rho$-exchanging t-channel term is added in this work, and its contribution is increasing as $E_{\rm cm}$ grows.
Thus, the contact term dominates the amplitude near $\pi N$ threshold, and the $\rho$-exchanging term takes more contribution at larger $E_{\rm cm}$.

We give $E_{0+}$ with $g_{\rho NN}=0$ in Fig. \ref{multipolerho}, and one can estimate the $\rho$ meson contribution by comparing with full $E_{0+}$. From the figure, the $\rho$ exchanges are indispensable in both $I_3=\frac{1}{2}$ and $I_3=-\frac{1}{2}$ channels around 1200$\sim$1400 MeV.

Here, the $\omega$-exchanging channel is not taken into account, since it behaves similarly like the $\rho$-exchanging one and thus can be combined into $\rho$-exchanging contribution. The related coupling parameters about the $\omega$ meson are also set 0 in Ref. \cite{Matsuyama:2006rp}.

The main part of amplitude in Eq.  (\ref{M0_piN}), the contact term, is with isospin factor $\tau_3$ being opposite for $I_3=\frac{1}{2}$ and $I_3=-\frac{1}{2}$. This makes the electric dipole $E_{0+}$ amplitude having opposite signs between $p$ and $n$ target around $\pi N$ threshold.

Similar trends of $\hat{\mathcal{M}}_{\eta N}^{(0)}$ for each channel are retained, but begins from $\eta N$ threshold. The contact term is discarded  in $\hat{\mathcal{M}}_{\eta N}^{(0)}$ due to flavor forbidden, and hence the magnitude of $\hat{\mathcal{M}}_{\eta N}^{(0)}$ contribution is minor to that of $\hat{\mathcal{M}}_{\pi N}^{(0)}$.

There are two terms in Eq. (\ref{t-tot}). The first one $V^{\lambda_\gamma, \lambda_N}_{\pi N,\gamma N}$ is directly from the $\hat{\mathcal{M}}_{\pi N}^{(0)}$ and contributes to more than 80\% of ${\rm Re}(E_{0+})$ up to 1400 MeV in our framework. The second term receives the corrections from the hadron rescattering $T_{\pi N,\alpha}$ and produces the imaginary part of $E_{0+}$.

When considering the contributions of $N^*(1535)$, we first use the bare propagator of the resonance to obtain the pure electromagnetic potentials, and the FSI effects recover its physical mass and generate the decay width naturally. As the energy approaches the bare mass $m_{N^*}^0$, the two terms in Eq. (\ref{t-tot}) are both divergent due to the lack of a finite imaginary part in the bare propagator. However, the two divergences cancel each other if the couplings $g_{\pi N N^*}$ and $g_{\eta N N^*}$ are the same for the pure electromagnetic potentials and FSI effects. This subtle cancellation indicates the consistence of different parts in our framework.

Another simple treatment for the $N^*(1535)$ contribution can be that the resonance propagator uses the Breit-Wigner form depending on the physical mass and width without FSI effects. Some dynamics would be missed in that way.

If $N(1535)$ is purely dynamically generated by the $\pi N$, $\eta N$ interactions and so on without the bare triquark kernel, $\hat g_{N\gamma N^*}$ would be 0 and the $N^*(1535)$ contribution would only appear in the FSI effects. We provide $E_{0+}$ with $\hat g_{N\gamma N^*}=0$ in Fig. \ref{multipolerho}, and one can clearly see that $\hat g_{N\gamma N^*}$ term is important around 1.5 GeV. The bare triquark kernel of $N^*(1535)$ makes the real part of $E_{0+}$ on $p$ target lower and imaginary part larger, which improves the consistence between our results and the empirical data, especially for the imaginary parts. Similar improvement can be found in the $n$ target case though not very obvious as in the $p$ target case.

While the present description are well agreed with SAID electric dipoles, there are some residual  discrepancy near 1.6 GeV that could be traced back to the absence of $N^*(1650)$ resonance and its interference, and $K\Lambda$, and $K\Sigma$ coupled-channel contribution as mentioned in Refs. \cite{Doring:2009uc,Doring:2013glu,Doring:2009yv}. Also the $\pi\pi N$ contribution may improve the consistence better at low energies.

Now we focus on the differences from other studies of photoproduction processes and therein one can notice the improvements in this work.
The partial-wave analysis, K-matrix methods and their variations are widely used to reproduce large amounts of experimental data from photoproduction and pion production processes \cite{Arndt:2002xv,Inoue:2001ip,Shklyar:2004ba,Penner:2002md,Usov:2005wy,Shyam:2008fr,Drechsel:1998hk,Drechsel:2007if,Tiator:2011pw,Workman:2012jf}.
When implementing the K-matrix approximation, the real parts of meson-baryon propagator in coupled-channel integral equations are omitted for technical simplicity, and thus the integral equations are reduced to algebraic equations. Such methods sometimes cannot show the essential dynamics.

The chiral perturbation theory (CHPT) can give a nice description of photoproduction observables near threshold where the spontaneously chiral symmetry breaking plays a central role \cite{Ma:2020hpe,GuerreroNavarro:2019fqb,Bernard:1991rt,Bernard:1994gm,Bernard:2007zu,Hilt:2013uf}.
CHPT can improve its results order by order.
But when moving to the resonance region, one is still facing some challenges, for example, the convergence may not be good. Comparing with relevant ChPT works, our results also give a good consistence at low energies though not as precise as Refs. \cite{Ma:2020hpe,Bernard:1994gm} and can extend well to the larger energies.

Dynamical coupled-channel models and some equivalent models are used to analyze the extensive and precise meson- or photon-production data and then extract the resonance parameters \cite{Matsuyama:2006rp,Kamano:2013iva,Kamano:2016bgm,Doring:2009uc,Doring:2013glu,Doring:2009yv,Pascalutsa:2004pk,Fuda:2003pd,Chen:2007cy}.
In these models, plenty of resonances and coupled channels are considered, where exhaustedly include the almost all experimental data, such as total cross section, differential cross sections, polarization observables, and helicity amplitudes.
As a consequence, though with more disagreement appearing in some specific partial-wave amplitude or polarization observables due to huge amounts of data taken into account, these models provide good interpretations to the experimental data and complement to partial-wave analysis group SAID \cite{gws}.
Apparently, these models often aim at describing the raw experimental data as precise as possible, determining the partial-wave amplitude, then extracting the resonances parameters like mass, widths, quantum numbers, couplings, magnetic moments, helicity amplitudes, and transition form factors, and so on.

\begin{figure}
        \setcounter{figure}{2}
        \centering
        \begin{tabular}{cc}
        \includegraphics[height=5cm]{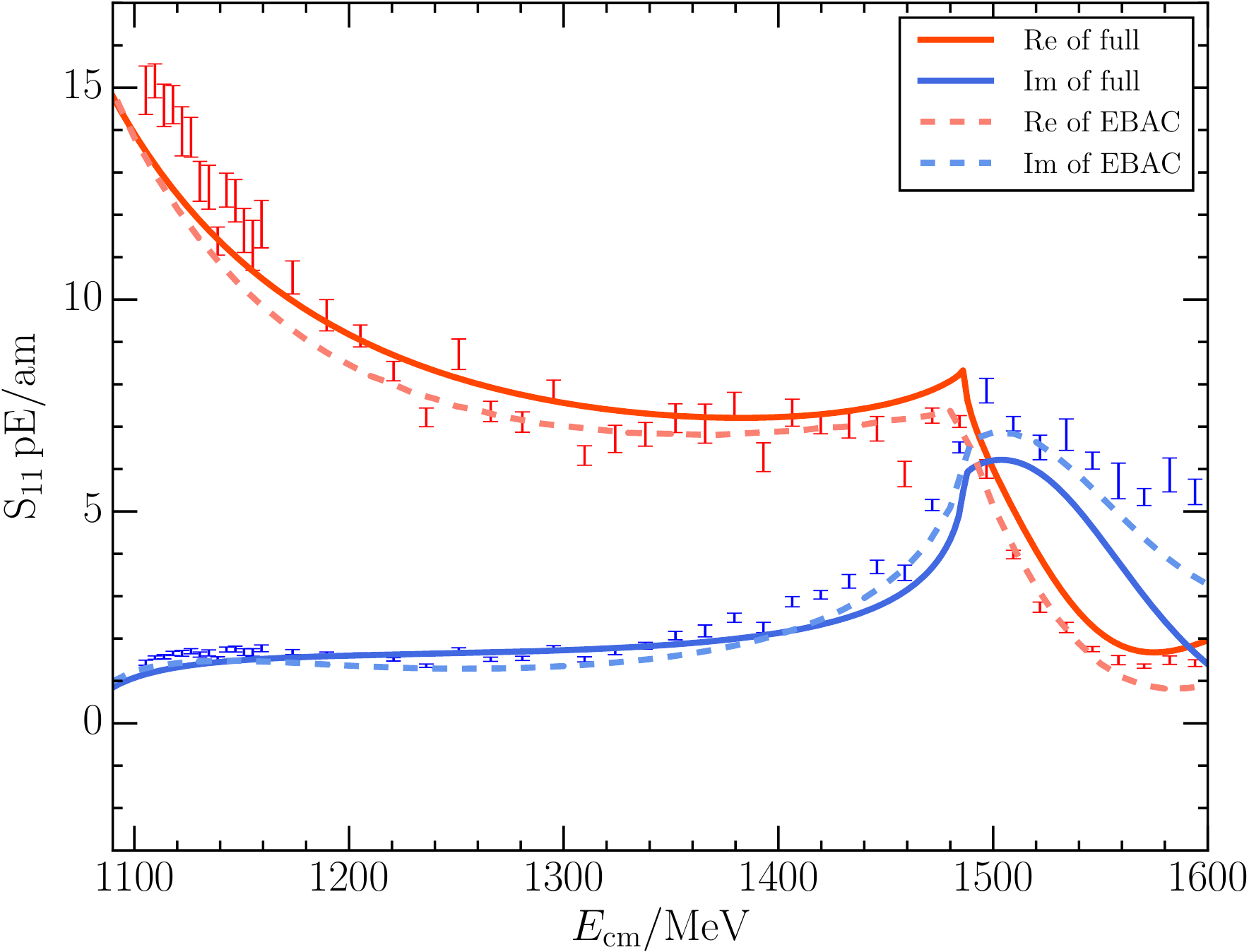}&  

        \end{tabular}
        \caption{The electric dipole amplitudes $E_{0+}$ from the SAID \cite{gws}, EBAC \cite{DCC model}, and our results. }
        \label{multipoleEbac}
      \end{figure}

The well-performed dynamical coupled-channel model, EBAC \cite{Matsuyama:2006rp,Kamano:2013iva,Kamano:2016bgm}, performing a dynamical coupled-channel analysis of worldwide data about total cross section, differential cross sections, polarization observables from pion- and photon-induced reactions, also provides the electromagnetic multipoles online \cite{DCC model}. We plot the EBAC, SAID and our results for $E_{0+}$ on proton target in Fig. \ref{multipoleEbac}, and noticed that EBAC and SAID extractions are consistent below 1600 MeV. From the figure, our real (imaginary) part is closer to the EABC data near threshold (at larger energies).

A dynamical coupled channel framework and semiphenomenological approach \cite{Huang:2011as,Ronchen:2014cna}, both with the FSI effects provided by the $\mathrm{J\ddot{u}lich}$ model, have studied the cross section, differential cross section, polarization observable and helicity amplitude plots for the photoproduction processes at different c.m. energies, and do exhibit a well description.
The analyticity is guaranteed in Ref. \cite{Ronchen:2014cna} and the $\chi^2$ of $pE_{0+}$ is 1574 (${\rm am}^2$) there and smaller than 2136 in this work below 1600 MeV.
Reference \cite{Huang:2011as} also suggests that the fitting results of multipole amplitude is very difficult and uncertain.

We do not include extra form factors for the electromagnetic vertices since the energy does not change too widely, which can largely reduce the manmade arbitrariness and uncertainties. However, the adjustable cutoffs in the electromagnetic form factors are critical for the photoproduction studies with most dynamical coupled-channel models \cite{Huang:2011as,Ronchen:2014cna}.

Another interesting discovery, we find the $\rho$-exchanging t-channel term is generally considered in Refs. \cite{Huang:2011as,Ronchen:2014cna,Doring:2009uc,Matsuyama:2006rp,Kamano:2013iva,Kamano:2016bgm}, but is not discussed very much.
However, as seen in Fig. \ref{multipolerho}, the $\rho$ exchanging channel plays an important role in this work to regulate the magnitudes of $p$ and $n$ target multipoles, simultaneously making the modulus of $p$ target higher and $n$ target lower around 1.5 GeV, since the contributions of $\rho$ exchanging have same signs for the $I_3=\frac{1}{2}$ and $I_3=-\frac{1}{2}$ processes.
The $\rho$ exchange provide the medium and short range force, which makes it become bigger around 1.5 GeV than threshold. However, the role of $\rho$ exchange in the interactions among the photon, meson, and nucleon need be checked in other works and may be more obvious in other channels.

To explain the mass inverse problem between $N^*(1535)$ and $N^*(1440)$ and sizable couplings of $N^*(1535)$ to $K\Lambda$, $N\eta'$ and $N\phi$ channels, it is proposed that the $N^*$ resonances contain 5-quark components  \cite{Zou:2010tc,An:2008xk,Liu:2005pm}.
The interference of $N^*(1535)$ from $N^*(1650)$ are significant, and strong couplings to $K\Lambda$ and $K\Sigma$ also suggest dynamical generation nature \cite{Doring:2009uc}.
Reference \cite{Inoue:2001ip,Kaiser:1995cy,Doring:2009uc} discuss the dynamical generation mechanism of $N^*(1535)$, generated from the pure meson and baryon interactions, but the predicted amplitudes appear not to be in good agreement with experiment.
The $N^*(1535)$ structure is still a well-known unsolved question.
From Fig. \ref{multipolerho}, the imaginary part of the S11 multipole deviates from the experimental data a little bigger than the real part starting at $E_{cm}\sim$1450 MeV. This may be improved by including the contributions from $\pi \pi N$, $K\Lambda$ $K\Sigma$, $N^*(1650)$, and their interference, which would disclose the fine structure of the $N^*(1535)$.

These dynamical coupled-channel models can extract the parameters for the resonances and help us understand the structure, but hadron structure models or lattice QCD calculations can provide other useful information as mentioned in \cite{Crede:2013kia,Burkert:2004sk}.
Lattice QCD starts from the first principles of QCD and is a model-independent approach.
Usually it gives the results depending on unphysical quark masses with finite volume effects.
Thus a comprehensive analysis, interpreting the results of lattice QCD simulation, meson-nucleon scattering, and photon-induced production, is appealing.

In our previous work $N^*(1535)$ is interpreted as a primary three-quark state, meanwhile with some five-quark components  \cite{Liu:2015ktc}.
It contains about 50\% of the bare triquark part near physical pion mass, within lattice boxes $L\simeq$ 2, 3 fm, by comparing the predicted energy levels of HEFT and the positions of finite-volume lattice QCD simulations. Moreover, as analytically continuing the scattering amplitude to complex plane, the resonance pole locating on the unphysical sheet are discovered as 1531$\pm$29-88$i\pm$2 MeV, which is consistent with the PDG value \cite{ParticleDataGroup:2020ssz}.

Looking to the bare electromagnetic coupling constants, they are fitted as $g_{p\gamma N^{*+}}=0.27$ and $g_{n\gamma N^{*0}}=-0.25$, and originate from the contribution of the triquark component in $N^*(1535)$.
As shown in Fig. \ref{multipolerho}, a nonvanishing bare $\hat{g}_{N\gamma N^*}$ must exist.
If comparing the physical electromagnetic couplings extracted from the electromagnetic decay widths of $N(1535)$ in PDG \cite{ParticleDataGroup:2020ssz}, $|g^{\rm phy}_{p\gamma N^{*+}}|=0.613^{+0.095}_{-0.112}$ and $|g^{\rm phy}_{n\gamma N^{*0}}|=0.466^{+0.180}_{-0.337}$, one can see that the bare couplings are not small.
The bare state cannot be absent in this work before considering more coupled channels and resonances.

\begin{figure}
        \centering
        \begin{tabular}{c}
        \includegraphics[height=5cm]{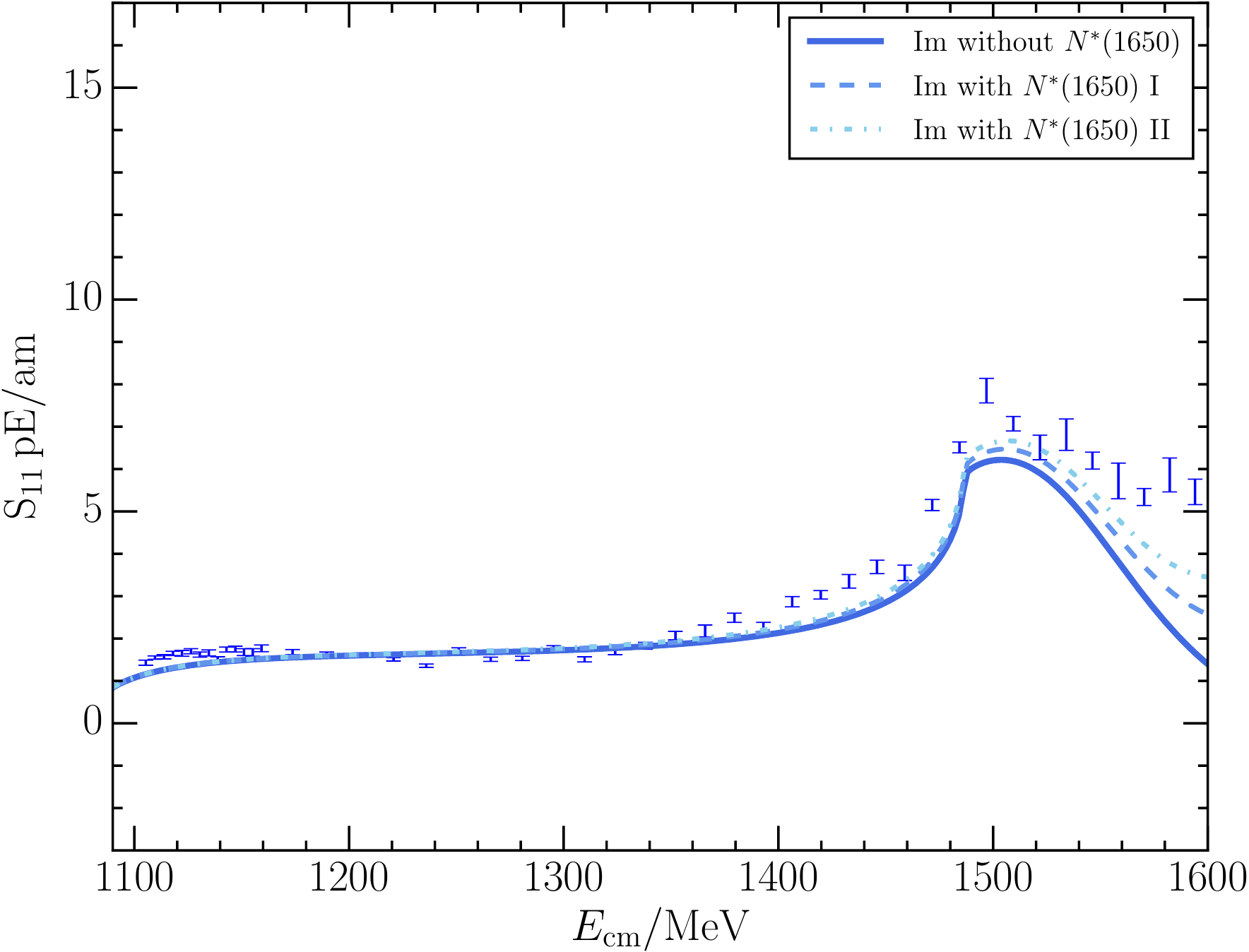}
        \end{tabular}
        \caption{The imaginary electric dipole amplitudes $E_{0+}$ on the proton target with the consideration of $N^*(1650)$. }
        \label{multipoleN1650}
      \end{figure}

In Ref. \cite{Doring:2009uc}, the simultaneous study of the $\pi N\to \pi N$ and $\gamma N \to \pi N$ reactions examines the structure of $N^*(1535)$ and the interference effect from $N^*(1650)$. The poles of $N^*(1535)$ and $N^*(1650)$ can be found on different Riemann sheets intuitively and their roles are carefully analyzed. The interference of $N^*(1650)$ is important for the properties of $N^*(1535)$ therein.

The biggest deviations of our results from the SAID data are the imaginary parts of the proton target. $K\Lambda$ and $K\Sigma$ thresholds are larger and thus they may contribute little to the imaginary parts below 1.6 GeV. We would focus on the estimation of $N^*(1650)$ resonance for the imaginary parts.

It is natural to expect the inclusion of $N^*(1650)$ with a sizable width will promote the present status of the imaginary amplitude near 1.6 GeV.
      For a rough estimation, we add a Breit-Wigner form for the $N^*(1650)$ contribution with physical mass and width from PDG, and explicitly,
      \begin{eqnarray}\label{eq:M_Nstar}
        &&\hat{\mathcal{M}}^{(N^*(1650))}= e \frac{\sqrt{3} g^{\rm phy}_{\pi N N^*(1650)}}{f_\pi} \frac{\hat{g}^{\rm phy}_{N\gamma N^*(1650)}}{4m_N}\nonumber \\&&
        \left\{ \slashed{k} \frac{\slashed{q}+\slashed{p}+m_{N^*(1650)}}{(\slashed{q}+\slashed{p})^2-(m_{N^*(1650)})^2+
        im_{N^*(1650)} \Gamma_{N^*(1650)}}  \gamma_5(\slashed{q}\slashed{\varepsilon}_\gamma- \slashed{\varepsilon}_\gamma\slashed{q})
        \right\} . \qquad\nonumber
      \end{eqnarray} 
Then using the Eqs. (\ref{explicitAmp})-(\ref{JLSheliPoten}) we can obtain $V_{\pi N,\gamma N}^{\lambda_\gamma=1, \lambda_N=\frac12,(N^*(1650))}(k,q)$, where the FSI contributions are included in the width of $N^*(1650)$ effectively.

From central values of PDG decay information, we can obtain $|g^{\rm phy}_{p\gamma N^{*+}(1650)}|=0.336$, $|g^{\rm phy}_{n\gamma N^{*+}(1650)}|=0.285$ and $g^{\rm phy}_{\pi N N^{*}(1650)}=0.092$. We plot the imaginary electric dipole amplitudes $E_{0+}$ on the proton target with these couplings for the effect of $N^*(1650)$ in Fig. \ref{multipoleN1650} as the Scenario I. The imaginary amplitude is promoted about 0.50 am at $E_{cm}=$1550 MeV. When taking the couplings from the upper limit of PDG widths, the scenario II in Fig. \ref{multipoleN1650} shows a growth of 0.87 am at $E_{cm}=$1550 MeV.

Unfortunately, our photoproduction amplitudes are no longer gauge invariant for either $\hat{\mathcal{M}}_{\alpha}^{(0)}$ or $\hat{\mathcal{M}}_{\alpha}^{(N^*)}$ after taking into account the final state interaction.
      One can refer to Refs. \cite{Haberzettl:2011zr,Haberzettl:2006bn} for the discussion about gauge invariance and off-shell condition when taking account of final-state interaction.
        The guiding principles of construction of gauge invariant amplitude are the consistent and complete implementation to all levels of reaction mechanisms \cite{Haberzettl:2011zr,Haberzettl:2006bn}.
      In the processes $\gamma p\to\pi^0\eta p$ and $\gamma p\to\pi^0K^0\Sigma^+$, when considering the FSI effect and using Bethe-Salpeter equation with the kernel of the Weinberg-Tomozawa term, the gauge invariance is automatically satisfied as long as the photon not only attached to external legs and vertexes, but also to intermediate propagators and vertexes \cite{Doring:2005bx}.
      We should also consider these effects in the future.

In short, a consistent understanding of $N^*(1535)$ is obtained with simultaneously interpreting QCD simulation, meson-nucleon scattering amplitudes and photon-induced dipole amplitudes.

\section{Summary}\label{summary}

The photoproduction information of nucleon resonance is particularly interesting to reflect the internal structure, where both hadronic and electromagnetic coupling are accessible.
Electromagnetic properties afford some additional information for the structure of strong interaction systems. The multipoles are directly tied to the scattering amplitudes and one can gain the insight into the nature of resonances more sensitively.

We analyze the role of $N^*(1535)$ and hadron interactions in the $\gamma N\to\pi N$ and $\pi N\to\pi N$ reactions and lattice QCD simulation simultaneously together with our previous work \cite{Liu:2015ktc}. We extend HEFT for the $\gamma N\to\pi N$ process, and the parameters for the FSI part are the same as previous ones. First we give the general $\gamma N\to\pi N$ potentials without FSI and transform them among different representations, and then provide the corresponding scattering T matrix with FSI and associate electric dipole amplitudes.

From the numerical results, the contact term dominates the region near the threshold, and the $\rho$-exchanging contributions become important around 1200$\sim$1400 MeV.
The $N^*(1535)$ resonance is dynamically generated among the bare triquark kernel and $\pi N$ and $\eta N$ channels in our framework \cite{Liu:2015ktc}. The coupling of the bare $N^*(1535)$ with $\gamma N$ is indispensable for the electric dipole amplitudes around 1500 MeV.

Combined with out previous work \cite{Liu:2015ktc}, an excellent picture of simultaneous descriptions is revealed for pion photoproduction, $\pi N$ scattering, resonance pole position, and the finite-volume energy level and structure analysis associated with lattice QCD simulations, in spite of certain residual discrepancy in the end of considered energy range, near 1.6 GeV. The $S_{11}$ partial wave meson-nucleon channel with negative parity is complicated, since there are many open thresholds such as $K\Lambda$ and $K\Sigma$ and also two interfering resonances $N^*(1535)$ and $N^*(1650)$. Including these effects sequentially in future will provide a better deciphering of nucleon resonances and understanding of photo- and hadroinduced production.

\section{Acknowledgment}
We would like to thank Dr. Hiroyuki Kamano for useful discussion. This project is supported by the National Natural Science Foundation of China under Grants No. 12175091, No. 11965016, and No. 12047501, and CAS Interdisciplinary Innovation Team.


\end{document}